\documentclass[sigconf]{acmart}

\usepackage{booktabs} 
\usepackage{multirow}
\usepackage{graphics}
\usepackage{graphicx}
\usepackage{algorithm}
\usepackage{algorithmic}
\usepackage{amsfonts}
\usepackage{amsmath}
\usepackage{color, url}
\usepackage[skip=0pt]{caption}
\usepackage[skip=0pt]{subcaption}
\usepackage{tabto}
\usepackage{url}
\usepackage{xcolor, colortbl}
\usepackage{booktabs}
\usepackage{enumitem}
\usepackage{bbm}
\usepackage{dsfont}
\usepackage{soul}

\usepackage{bm}

\usepackage{url}

\usepackage{subcaption}

\usepackage[mathscr]{eucal}

\usepackage{bigstrut,multirow,rotating}

\usepackage{placeins}

\graphicspath{ {figs/} }
\usepackage{algorithm,algorithmic}

\usepackage{balance}
\usepackage{makecell}

\newcommand{\myparatight}[1]{\smallskip\noindent{\bf {#1}:}~}

\copyrightyear{2022}
\acmYear{2022}
\setcopyright{acmcopyright}
\acmConference[SACMAT '22] {Proceedings of the 27th ACM Symposium on Access Control Models and Technologies}{June 8--10, 2022}{New York, NY, USA.}
\acmBooktitle{Proceedings of the 27th ACM Symposium on Access Control Models and Technologies (SACMAT '22), June 8--10, 2022, New York, NY, USA}
\acmPrice{15.00}
\acmISBN{978-1-4503-9357-7/22/06}
\acmDOI{10.1145/3532105.3535023}

\begin{document}

\fancyhead{}

\newcommand{\kliu}[1]{{\color{blue}{\bf Kevin:} #1}}
\newcommand{\kliuc}[1]{{\color{blue} #1}}

\title{FairRoad: Achieving Fairness for Recommender Systems with Optimized Antidote Data}

\author{Minghong Fang}
\affiliation{%
	\institution{The Ohio State University}
}

\author{Jia Liu$^{*}$}
\affiliation{%
	\institution{The Ohio State University}
}

\author{Michinari Momma$^{*}$}
\affiliation{%
	\institution{Amazon}
}

\author{Yi Sun$^{*}$}
\affiliation{%
	\institution{Amazon}
}

\thanks{$^*$This work does not relate to positions at Amazon}

\begin{abstract}
	Today, recommender systems have played an increasingly important role in shaping our experiences of digital environments and social interactions.
	However, as recommender systems become ubiquitous in our society, recent years have also witnessed significant fairness concerns for recommender systems.
	Specifically, studies have shown that recommender systems may inherit or even amplify biases from historical data, and as a result, provide unfair recommendations.
	To address fairness risks in recommender systems, 
	most of the previous approaches to date are focused on modifying either the existing training data samples or the deployed recommender algorithms, but unfortunately with limited degrees of success.
	In this paper, we propose a new approach called \ul{fair} \ul{r}ecommendation with \ul{o}ptimized \ul{a}ntidote \ul{d}ata (FairRoad), which aims to improve the fairness performances of recommender systems through the construction of a small and carefully crafted antidote dataset. 
	Toward this end, we formulate our antidote data generation task as a mathematical optimization problem, which minimizes the unfairness of the targeted recommender systems while not disrupting the deployed recommendation algorithms. 
	Extensive experiments show that our proposed antidote data generation algorithm significantly improve the fairness of recommender systems with a small amounts of antidote data.
\end{abstract}

\ccsdesc[500]{Information systems~Recommender systems}

\keywords{Fairness, Recommender Systems, Antidote Data.}

\maketitle


\section{Introduction} \label{sec:intro}

{\bf Background and Motivation:} Today, we are interacting with a large variety of recommender systems on a daily basis, including but not limited to finding accommodation (Airbnb and Booking.com), online video streaming (YouTube and Dailymotion), online music streaming (Spotify, Pandora, and Soundcloud), ride-sharing (Uber and Lyft), online shopping (eBay), Apps searching (Apple and Google App Store), job searching (LinkedIn and Indeed), and even in some high-risk decision making settings, such as financial loan approval, hiring, etc. \cite{Metevier19:Contextural_Bandit_Fairness}.
of digital environments and social interactions\cite{Milano20:RecommenderSystem}.
By analyzing users' historical data and interactions (e.g., ratings and reviews), recommender systems have fundamentally changed our social life by facilitating efficient information retrieval and context-aware searching and ranking results. 
As such, recommender systems have played an increasingly important role in shaping our experiences of digital environments and social interactions.

However, despite all the above salient features of recommender systems, recent studies have found that recommender systems could provide {\em unfair} and/or {\em biased} recommendations to users if not designed and maintained appropriately~\cite{yao2017beyond,zhu2020measuring}. 
For instance, an investigation on more than one million ratings in MovieLens dataset \cite{MovieLens} has showed that, over time, the algorithm amplified the initial bias toward males and recommendations are deviated from female users' preferences.
Moreover, as a result of the deviation, the dataset is gradually dominated by ratings from males and losing its female population.
As another example, it has been reported \cite{FacebookAds,Ali19:FacebookAds} that job ads on Facebook for supermarket cashier positions were shown to 85\% women, and job positions in Taxi companies were shown to audience that was almost 75\% black.

To date, there have been some existing works attempting to address the fairness risks in recommender systems~\cite{islam2021debiasing,steck2018calibrated,yao2017beyond,garcia2021maxmin,patro2020fairrec,zhu2021popularity,burke2018balanced,zhu2020measuring,wang2021deconfounded,gao2020counteracting,wu2021learning}.
For instance, the authors in~\cite{islam2021debiasing} proposed a pre-training and fine-tuning method to reduce the gender bias in career recommendations without sacrificing the performance of the learnt system. 
Yao et. al~\cite{yao2017beyond} proposed a regularization-based approach to improve the fairness of recommender systems by incorporating an unfairness term into the objective function.
The authors in ~\cite{steck2018calibrated} propose a re-ranking algorithm to post-process the recommendation list to achieve fairness.
However, these existing works are ``intrusive'' in the sense that they need to modify either the original training data or the learning algorithms of the recommender systems, both of which are quite cumbersome in practice.
These limitations motivate us to pursue a different approach in this paper.

{\bf Our work:} In this paper, we propose a new approach named {\em FairRoad} (\ul{fair} \ul{r}ecommendation with \ul{o}ptimized \ul{a}ntidote \ul{d}ata) to achieve fairness for recommender systems without modifying the original data nor recommendation algorithms.
The {\em basic idea} of our proposed FairRoad method is that we assume the service provider (e.g., the companies who provide recommendations to users) is able to inject $m$ antidote users into the recommender systems.
Each antidote user can at most give rating scores to $n$ items (called \textit{filler items}).
Clearly, by injecting an appropriate amount of antidote users with carefully crafted ratings for filler items, the learning outcome of the recommender systems will be changed, so that fairness risks of the recommender systems can be mitigated.
Toward this end, a key question naturally arises: {\em given some targeted unfairness metric (more on this later), how to optimally select filler items for each antidote user and assign rating scores to those items, such that the unfairness score is minimized after training the model on the original and injected antidote data?}

To answer this fundamental question, we formulate our antidote data generation problem as a mathematical optimization problem, for which the objective function is to minimize some targeted unfairness metric.
However, it turns out that the formulated optimization problem for antidote data generation is a highly challenging integer programming problem, which is computationally intractable.
To address this challenge, we propose several effective techniques to lower the complexity of the formulated optimization problem.
First, we convert the rating scores of integer values to continuous variables, solve the relaxed continuous optimization problem, and then project the solution back to integer rating scores.
Second, instead of computing the rating scores of $m$ antidote users all at once, we optimize rating scores of antidote users one by one in a sequential fashion.
Specifically, in each iteration, we add one antidote user into the recommender system and compute the optimized rating scores for this antidote user.

A major novelty of this paper is that, upon completing the algorithmic design of antidote data for a single unfairness performance metric, we go one step further to consider optimizing antidote data for optimizing or balancing {\em multiple} unfairness metrics.
This is motivated by the observations from recent studies~\cite{patro2020fairrec,binns2020apparent,berk2021fairness} that improving one unfairness metric for recommender systems may often worsen the performance of another unfairness metric.
That is, different unfairness metrics may conflict with one another.
So far, most of the existing works in the literature are focused on how to reduce the fairness risks in terms of a single unfairness metric.
However, how to simultaneously reduce the fairness risks of multiple unfairness metrics remain largely unexplored.
Note that this multi-unfairness antidote data optimization problem is in essence a multi-objective optimization problem.
In the literature of multi-objective optimization, a classical problem formulation is to optimize the weighted sum of multiple unfairness metrics.
However, in practice, it is often hard to specify the weight for each unfairness performance metric.
Moreover, even if the weights for different unfairness metrics are known, simply using the gradient descent algorithm that follows the gradient directions of the weighted sum of multiple unfairness metrics is often inefficient.
This is due to the fact that most of the unfairness metrics of practical interest have highly complex geometric characteristics (convexity, curvature, etc.).

To address the above challenges, in our paper, we propose a gradient-deflecting approach motivated by \cite{yu2020gradient,wang2020gradient} to improve the performance of multi-unfairness antidote data generation.
Specifically, given two unfairness performance metrics that the service provider wants to optimize, we first compute the gradients of the objective functions of these two unfairness metrics.
Then, we examine whether gradients are conflicting by computing the cosine similarity between gradients.
If the cosine similarity is positive, which indicates the two gradients are not in conflict, we then sum the two gradients and a leverage projected gradient descent approach to generate rating scores of antidote users.
If on the other hand the cosine similarity is negative (i.e., gradients are in conflict), we modify each gradient by projecting one gradient onto the normal plane of the other one, which helps avoid being trapped at some point that has high curvature.
Finally, the two altered gradients are added to obtain the deflected gradient, which will guide the antidote data generation process.

We evaluate our FairRoad antidote data generation method on four synthetic and one real-world datasets.
We show that our FairRoad approach could improve the fairness of recommender systems with a small amount of antidote data.
Our FairRoad could reduce 31.25\% of overestimation unfairness when only injecting 2\% of antidote users into the synthetic dataset; 19.05\% of value unfairness is reduced with 2\% of antidote users on Yelp dataset.
Lastly, we summarize our main contributions in this paper as follows:

\begin{list}{\labelitemi}{\leftmargin=2em \itemindent=-0.0em \itemsep=.2em}
	\item 
	We propose a new method called FairRoad based on antidote data optimization to improve fairness for recommender systems.

	\item 
	Based on the algorithmic framework of FairRoad for a single unfairness performance metric, we further propose a new approach to improve the performance of multi-unfairness antidote data generation for recommender systems.
	
	\item 
	Extensive experiments on four synthetic and one real-world benchmark demonstrate that our proposed FairRoad algorithms are effective in mitigating the unfairness in recommender systems and outperform existing baselines.

\end{list}


\vspace{-.1in}
\section{Preliminaries and Related Work} \label{sec:related}

In this section, we first provide a quick overview of the basics of recommender systems and their fairness metrics 
in Section~\ref{subsec:recsys} and Section~\ref{fairness_metrics_def} to familiarize readers with the necessary background. 
Then, in Section~\ref{subsec:related}, we introduce some related work on fairness,
thus putting our work in a comparative perspective.
Table~\ref{Main_notations} 
summaries the key notation used in this paper.

 \begin{table}[t!]
 	\centering
 	\addtolength{\tabcolsep}{-0.0pt}
 	\caption{Notation.}
 	\label{Main_notations}%
 	\begin{tabular}{l|lll}
 		\toprule
 		Notation &  Definition  \\
 		\hline
 		$\mathcal{U}$   & Set of users \\
 		$\mathcal{I}$   & Set of items \\
 		$\left| \mathcal{U} \right|$  & Number of users \\
 		$\left| \mathcal{I} \right|$  & Number of items \\
 		$\mathcal{D} $   & Set of disadvantaged users\\
 		$\mathcal{A}$   & Set of advantaged users\\
 		$\mathcal{U}_i$  & Set of users who rate item $i$  \\
 		$\mathcal{I}_u$  & Set of items rated by user $u$  \\
 		$\mathcal{D}_i $  & Set of disadvantaged users who rate item $i$  \\
 		$\mathcal{A}_i$  & Set of advantaged users who rate item $i$  \\
 		\bottomrule
 	\end{tabular}%
 \end{table}%

\subsection{Basics of Recommender Systems} \label{subsec:recsys}
In this paper, we focus on recommender systems based on collaborative filtering using matrix factorization~\cite{koren2009matrix}, which is one of the most popular approaches for implementing recommender systems.
Toward this end, we use $\mathcal{U}$ and $\mathcal{I}$ to represent the set of users and items, respectively.
We let $\left| \mathcal{U} \right|$ and  $\left| \mathcal{I} \right|$ represent the numbers of users and items, respectively.
Each user is associated with some identity (e.g., male or female).
Let $ \bm{R} \in {{\mathbb{R}}^{\left| \mathcal{U} \right| \times \left| \mathcal{I} \right|}}$ denote the observed rating matrix, and $\Omega$ be all observable entries in $\bm{R}$.
An entry $r_{ui}$ in matrix $\bm{R}$ means the rating score that user $u$ gives to item $i$, where $u \in \mathcal{U}$, $i \in \mathcal{I}$.
We further let $\bm{p}_u \in {\mathbb{R}^d}$ and $\bm{q}_i \in {\mathbb{R}^d}$ be the $d$-dimensional latent vectors of user $u$ and item $i$, respectively.
In matrix-factorization-based recommender systems, the objective function can be defined as follows:
\begin{align}
\label{orig_matrix_opti_problem}
\mathop {\arg \min }\limits_{{\bm{P}},{\bm{Q}}} {\sum\nolimits_{(u,i) \in \Omega} {\left( {{r_{ui}} - \bm{p}_u^\top{\bm{q}_i}} \right)} ^2} + \lambda \left(  \left\| \bm{P} \right\|_F^2 + \left\| \bm{Q} \right\|_F^2 \right),
\end{align}
where $\bm{P} = [\bm{p}_1,\ldots,\bm{p}_{\left| \mathcal{U} \right|}]$, $\bm{Q} = [\bm{q}_1,\ldots,\bm{q}_{\left| \mathcal{I} \right|}]$, $\left\| \cdot \right\|_F$ is the Frobenius norm, and $\lambda$ is the regularization term.
The predicted rating of user $u$ gives to item $i$ can be computed as $\hat{r}_{ui} = \bm{p}_u^\top{\bm{q}_i}$, where $\bm{p}_u^\top$ is the transpose of vector $\bm{p}_u$.

\subsection{Fairness Metrics} \label{fairness_metrics_def}
To date, there have been several fairness metrics for recommender systems proposed in the machine learning community, including individual fairness~\cite{dwork2012fairness}, group fairness~\cite{calders2010three}, equalized odds~\cite{hardt2016equality} and demographic parity~\cite{calders2009building}. 
In this paper, we consider the four unfairness metrics that are applicable for matrix-factorization-based recommender systems for user groups that can be categorized into advantaged and disadvantaged users (e.g.,
the advantaged and disadvantage groups may be male and female users, respectively).

\smallskip
{\bf 1) Value Unfairness~\cite{yao2017beyond}:}
The value unfairness $ M_{\text{val}}$ metric measures the discrepancy in signed prediction errors between advantaged and disadvantaged users.
Value unfairness emerges when users in one group (advantaged or disadvantaged) always receive higher or lower predictions than their actual preferences. 
Specifically, the value unfairness $ M_{\text{val}}$ can be computed as follows:
\begin{align}
\label{Value_unfairness}
\!\!\!\!\! M_{\text{val}} = \frac{1}{\left| \mathcal{I} \right|} \sum\nolimits_{i \in \mathcal{I}} \left| \left(\mathbb{E}_\mathcal{D} [\hat{r}]_i - \mathbb{E}_\mathcal{D} [r]_i \right) -  \left(\mathbb{E}_\mathcal{A} [\hat{r}]_i - \mathbb{E}_\mathcal{A} [r]_i \right) \right|, \!
\end{align}
where $\mathcal{D}$ and $\mathcal{A}$ denote the sets of disadvantaged and advantaged users, respectively; 
$\mathbb{E}_\mathcal{D} [\hat{r}]_i$ is the average predicted rating score for item $i$ from disadvantaged users, 
$\mathbb{E}_\mathcal{A} [\hat{r}]_i$ is the average predicted rating score for item $i$ from advantaged users, and $\mathbb{E}_\mathcal{D} [{r}]_i$ and  $\mathbb{E}_\mathcal{A} [{r}]_i$ are the average rating scores for item $i$ from the disadvantaged and advantaged users, respectively.
More specifically, $\mathbb{E}_\mathcal{D} [\hat{r}]_i$ can be calculated as:
\begin{align}
\mathbb{E}_\mathcal{D} [\hat{r}]_i = \frac{1}{\left| \mathcal{D}_i \right| } \sum\nolimits_{u \in \mathcal{D}_i } \hat{r}_{ui},
\end{align}
where $ \mathcal{D}_i$ is the set of disadvantaged users who rate the item $i$. $\mathbb{E}_\mathcal{A} [\hat{r}]_i$, $\mathbb{E}_\mathcal{D} [r]_i$  and $\mathbb{E}_\mathcal{A} [r]_i$ can be computed similarly.

\smallskip
{\bf 2) Absolute Unfairness~\cite{yao2017beyond}:} 
The absolute unfairness $ M_{\text{abs}}$ metric captures the discrepancy in absolute prediction errors between advantaged and disadvantaged users.
The absolute unfairness $ M_{\text{abs}}$ is large when users in one group have small estimation errors and users in the other group have large estimation errors.
The absolute unfairness $ M_{\text{abs}}$ can be computed as:

\begin{align}
\label{Absolute_unfairness}
\!\!\!\! M_{\text{abs}} = \frac{1}{\left| \mathcal{I} \right|} \sum\nolimits_{i \in \mathcal{I}} \left\|\mathbb{E}_\mathcal{D} [\hat{r}]_i - \mathbb{E}_\mathcal{D} [r]_i \right| -  \left|\mathbb{E}_\mathcal{A} [\hat{r}]_i - \mathbb{E}_\mathcal{A} [r]_i \right\|, \!
\end{align}

\smallskip
{\bf 3) Overestimation Unfairness~\cite{yao2017beyond}:} 
Overestimation unfairness $M_{\text{over}}$ measures the discrepancy in how much the predictions overestimation the true ratings.
$M_{\text{over}}$ is large when the systems constantly predict higher rating scores than the true rating scores.
Overestimation unfairness $M_{\text{over}}$ is computed as the following:
\begin{multline}
\label{Over_unfairness}
M_{\text{over}} = \frac{1}{\left| \mathcal{I} \right|} \sum\nolimits_{i \in \mathcal{I}} \left| \text{max} \left\{0,  \mathbb{E}_\mathcal{D} [\hat{r}]_i - \mathbb{E}_\mathcal{D} [r]_i \right\} \right. \\
\left. - \text{max} \left\{0, \mathbb{E}_\mathcal{A} [\hat{r}]_i - \mathbb{E}_\mathcal{A} [r]_i \right\}  \right|. 
\end{multline}

{\bf 4) Non-parity Unfairness~\cite{yao2017beyond}:} 
Non-parity unfairness $M_{\text{par}}$ captures the absolute unfairness of predictions for advantaged and disadvantaged users.
Non-parity unfairness $M_{\text{par}}$ is computed as the absolute difference between the average predicted rating scores of advantaged and disadvantaged users.
To be specific, $M_{\text{par}}$ can be computed as:
\begin{align}
\label{parity_unfairness}
M_{\text{par}} = \left| \mathbb{E}_\mathcal{D} [\hat{r}] - \mathbb{E}_\mathcal{A} [\hat{r}]  \right|,
\end{align} 
where $\mathbb{E}_\mathcal{D} [\hat{r}]$ and $\mathbb{E}_\mathcal{A} [\hat{r}] $ are the average predicted rating scores of disadvantaged and advantaged users, respectively.

\subsection{Related Work} \label{subsec:related}

Noting that recommender systems are a quintessential application of machine learning technologies, in this section, we provide a quick overview on the broader algorithmic fairness research for machine learning systems.
In recent years, algorithmic fairness research has received increasing attention in the learning community~\cite{hardt2016equality, kirnap2021estimation, garcia2021maxmin, zhang2021omnifair, patro2020fairrec, wu2021learning, zhu2021popularity, islam2021debiasing, speicher2018unified, gao2020counteracting, fu2020fairness, binns2020apparent, burke2018balanced, morik2020controlling, biswas2020machine, kim2020fact, zafar2017fairness, zhu2020measuring, kang2020inform,rastegarpanah2019fighting,petersen2021post,zafar2017parity,li2020towards,nandy2020achieving,chen2020bias,wu2021fairrec,jones2020selective,li2019fair,mansoury2019bias,bower2021individually,xiao2017fairness,wang2021fair,sharma2021fair,li2020fairness}.
Among these, one major line of research is focused on fairness in classification and regression, with a variety of fairness notions being proposed~\cite{dwork2012fairness,calders2010three,hardt2016equality,calders2009building,yao2017beyond,wan2020addressing,beutel2019fairness,zafar2017fairness}.
For example, individual fairness~\cite{dwork2012fairness} requires that two similar individuals should receive similar algorithmic outcomes.
Equal odds~\cite{hardt2016equality} requires that the true positive and false positive rates should be equivalent for different groups.
Disparate mistreatment~\cite{zafar2017fairness} ensures that classification error rate should be the same among different groups.
Inspired by the fairness notions from the machine learning community, Yao et. al~\cite{yao2017beyond} propose several unfairness metrics that can be used to matrix-factorization-based recommender systems, including value unfairness, absolute unfairness, overestimate unfairness and non-parity unfairness.
These unfairness metrics all measure the difference in estimation error of rating scores for disadvantaged and advantaged users.

Recently, there has also been another line of work focusing on mitigating the bias in machine learning and recommender systems.
According to different strategies that are used to improve fairness, these approaches fall into three categories: pre-processing, in-processing and post-processing.
Pre-processing techniques modify the (existing) input training data to reduce the recommendation bias, e.g., reweighting the input data~\cite{kamiran2012data} or learn fair representations of the original data~\cite{zemel2013learning}.
In-processing methods attempt to mitigate the discrimination during the model training process by, e.g., adding regularization penalty to the objective function~\cite{yao2017beyond}.
Different from pre-processing and in-processing methods that try to modify the existing input data or models to improve fairness, post-processing approaches achieve the desired fairness goal by changing the outputs of classifiers or recommender systems, e.g., re-ranking top-$N$ recommendation list~\cite{zehlike2017fa}.

Our antidote data framework has previously been considered as data poisoning attacks to machine learning models~\cite{fang2021data,cao2020fltrust,fang2020influence,fang2020local,biggio2012poisoning,chen2017targeted,jagielski2018manipulating,li2016data,alfeld2016data,shafahi2018poison,xiao2015feature,schwarzschild2021just,wu2021triple,wang2018data,jagielski2021subpopulation,koh2022stronger,cao2021data}.
The most related works to ours are~\cite{jang2021constructing,rastegarpanah2019fighting,solans2020poisoning}, where the proposed approaches all share some similarities to our antidote data approach. 
That is, all of these existing works attempt to affect the fairness of the model without directly modifying the original training data or the deployed learning algorithms.
In~\cite{jang2021constructing}, the authors trained a 
fair classifier on the synthetic fair dataset generated by the VAE-GAN based generative model to achieve fairness without sacrificing model's performance.
The authors in~\cite{rastegarpanah2019fighting} 
introduce an optimization-based scheme to inject extra data to
improve the polarization and fairness of recommender systems.
However, regardless of the purpose of the algorithms, 
all these existing work only focused on a single fairness metric.
In comparison, our FairRoad approach further consider the case of multiple unfairness metrics optimization.
As will be shown later, antidote data generation for multi-unfairness optimization in recommender systems is far from trivial and necessitates careful algorithmic designs.


\section{Problem Statement} \label{sec:problem}

In this section, we first formally present our proposed model, including the service provider's goal, capability, and knowledge in Section~\ref{subsec:sp_goal}.
We next mathematically formulate the service provider's goal as an optimization problem in Section~\ref{subsec:formulation}.

\subsection{Antidote Data Generation Model} \label{subsec:sp_goal}

\myparatight{1) Service provider's goal}
We consider a recommender system that is trained over user-item iteration data (e.g., rating scores) provided by users.
These interactions may be distributed unevenly among different user groups (e.g., advantaged and disadvantaged users) and typically exhibit bias. Recent studies have shown that recommendation model may inherit or even amplify the bias~\cite{zhu2020measuring,wang2021deconfounded}. 
The goal of the service provider is to improve the fairness of the targeted recommender system.
The service provider could be a company who provides web services to users.

\myparatight{2) Service provider's capability and knowledge}
In our FairRoad model, the service provider achieves his/her fairness goal by deliberately injecting some antidote data into the system such that the recommender system will make fair predictions when trained on the augmented dataset (original data plus the injected antidote data).
For instance, the employees in Netflix could inject a small number of well-crafted user-item interaction data into the database for social good, e.g., providing fair service to customers.

We assume that the service provider knows all details of the targeted recommender systems, including all the rating data, the parameters and algorithms used in recommendation model.
We note that, although appearing to be a very strong assumption, this setting is reasonable in practice since the rating data could be public and the service provider itself knows the details of the deployed recommender systems.

\subsection{Formulating Antidote Data Generation Model as an Optimization Problem} \label{subsec:formulation}

In this subsection, we formally formulate the antidote data generation model as an optimization problem to reduce the unfairness of the targeted recommender systems.
Suppose the rating scores of targeted recommender systems are restricted to be integers and can be chosen from the set $\left\{r_{\text{min}}, \cdots, r_{\text{max}} \right\}$, where $r_{\text{min}}$ and $r_{\text{max}}$ are the minimum and maximum rating scores, respectively.
We assume that the service provider could inject $\alpha \in [0,1)$ fraction of antidote users into the targeted recommender systems.
We let $\mathcal{U}$ and $\mathcal{\widetilde{U}}$ denote the set of original and antidote users, respectively.
Therefore, one has $\vert  \mathcal{\widetilde{U}} \vert = \lfloor \alpha \vert \mathcal{U}  \vert  \rfloor$.
Each antidote user will rate at most $n$ items, which are called filler items.
For an antidote user $z \in \mathcal{\widetilde{U}}$, we let $\bm{r}_z$ be the rating score vector of user $z$ and $\mathcal{I}_z$ be the set of items rated by $z$.
Then, we have $\vert \mathcal{I}_z \vert \le n$.
Let $r_{zi}$ be the rating score that antidote user $z$ gives to item $i$, thus one has $\bm{r}_z=[r_{zi}, i \in \mathcal{I}_z]^{\top}$.
Let $M_*$ be a given unfairness metric (e.g., value unfairness, absolute unfairness, overestimaiton unfairness, or non-parity unfairness).
The service provider aims to craft rating scores for each antidote user such that when the recommendation model is trained jointly on the original and antidote rating data, the unfairness metric $M_*$ is minimized.
Putting all modeling together, the unfairness metric minimization problem (UMM) can be formulated as the following:
\begin{align}
\label{objectiveFrame}
\text{ UMM:  min}  & \medspace\medspace \textstyle M_*  &  \\
\label{budgetconstraint}
\text{s.t. } \hspace{-0.25em} & \medspace\medspace r_{zi} \in \{r_{\text{min}},\cdots,r_{\text{max}}\},   && \forall z \in \mathcal{\widetilde{U}}, \forall i \in \mathcal{I}_z, &  \\
\label{itemSelected}
& \medspace\medspace \vert {\mathcal{I} _z} \vert  \le n,    && \forall z \in \mathcal{\widetilde{U}}, \hspace{0.25em} &  \\
& \medspace\medspace  \vert  \mathcal{\widetilde{U}} \vert = \lfloor \alpha \vert \mathcal{U}  \vert  \rfloor.
\end{align}

Note that the above formulation in Problem UMM is a general framework and can be applied to any recommender systems (matrix-factorization-based, association-rule-based, or graph-based) and unfairness metrics.
Given a targeted recommender system, solving the optimization problem for a given unfairness metric leads to a specific antidote data generation model for this particular unfairness metric.
However, we note that Problem UMM is a integer programming problem and is computationally intractable (the objective function involves the vector $\bm{r}_{z}$ and each value $r_{zi}$ in $\bm{r}_{z}$ is an integer variable).


\section{Our Solution} \label{attackModel}

In this section, we propose several efficient and practical techniques to solve the Problem UMM.
First, instead of optimizing the rating scores of all antidote users at the same time, we add one antidote user at a time to the targeted recommender system and optimize its rating scores in a sequential fashion.
However, even with sequential antidote user injection, it remains challenging to find the optimal rating scores for each antidote user since the rating score is an integer variable.
Inspired by~\cite{fang2020influence}, we relax the rating score to continuous variable, i.e., we let $\bm{x}_{z} = [x_{zi}, i \in \mathcal{I}_z]^{\top}$ be the relaxed rating score vector of antidote user $z$, where $x_{zi}$ is the continuous rating that injected antidote user $z$ gives to item $i$ and satisfies $x_{zi} \in \left[ r_{\text{min}}, r_{\text{max}}  \right]$.
We project the continuous rating score back to the integer values in the set $\left\{ r_{\text{min}}, \cdots, r_{\text{max}} \right\}$ after solving the optimization problem.
Based on the above relaxation, we reformulate Problem UMM as the following optimization problem for antidote user $z$:
\begin{align}
\label{appproblem}
\begin{split}
& \min_{\bm{x}_{z}} \mathcal{L}(\bm{x}_{z}) = M_* \\
& \text{   s.t. } x_{zi} \in [r_{\text{min}}, r_{\text{max}}].
\end{split}
\end{align}
In what follows, we will discuss in detail each component of our proposed solution approach.

\subsection{FairRoad: Antidote Data Generation for Single-Unfairness Optimization}

In this subsection, we will show how to solve the optimization problem in (\ref{appproblem}).
For illustrative purpose, we instantiate $M_*$ with the value unfairness metric $M_{\text{val}}$ introduced in Section~\ref{fairness_metrics_def}.
Then, the gradient of loss formulated in (\ref{appproblem}) can be computed as:
\begin{multline}
\label{sub_grad}
\bm{g}_{\text{val}} (\bm{x}_z) = \frac{1}{\left| \mathcal{I} \right|} \sum\nolimits_{i \in \mathcal{I}} \nabla_{\bm{x}_z} \left| \left(\mathbb{E}_\mathcal{D} [\hat{r}]_i - \mathbb{E}_\mathcal{D} [r]_i \right) 
\right. \\
\left. 
-  \left(\mathbb{E}_\mathcal{A} [\hat{r}]_i - \mathbb{E}_\mathcal{A} [r]_i \right) \right|.
\end{multline}
The term $\sum\nolimits_{i \in \mathcal{I}} \nabla_{\bm{x}_z} \left| \left(\mathbb{E}_\mathcal{D} [\hat{r}]_i - \mathbb{E}_\mathcal{D} [r]_i \right) 
-  \left(\mathbb{E}_\mathcal{A} [\hat{r}]_i - \mathbb{E}_\mathcal{A} [r]_i \right) \right|$ in (\ref{sub_grad}) can further be computed as:
\begin{align}
& \sum\nolimits_{i \in \mathcal{I}} \nabla_{\bm{x}_z} \left| \left(\mathbb{E}_\mathcal{D} [\hat{r}]_i - \mathbb{E}_\mathcal{D} [r]_i \right) 
-  \left(\mathbb{E}_\mathcal{A} [\hat{r}]_i - \mathbb{E}_\mathcal{A} [r]_i \right) \right| \nonumber \\
&=  \sum\nolimits_{i \in \mathcal{I}} \text{sign}   \left( \left(\mathbb{E}_\mathcal{D} [\hat{r}]_i - \mathbb{E}_\mathcal{D} [r]_i \right) 
-  \left(\mathbb{E}_\mathcal{A} [\hat{r}]_i - \mathbb{E}_\mathcal{A} [r]_i \right) \right)  \nonumber  \\
& \quad \cdot \nabla_{\bm{x}_z} \left(  \left(\mathbb{E}_\mathcal{D} [\hat{r}]_i - \mathbb{E}_\mathcal{D} [r]_i \right) 
-  \left(\mathbb{E}_\mathcal{A} [\hat{r}]_i - \mathbb{E}_\mathcal{A} [r]_i \right)   \right)  \nonumber  \\
&\stackrel{(a)} =  \sum\nolimits_{i \in \mathcal{I}} \text{sign}   \left( \left(\mathbb{E}_\mathcal{D} [\hat{r}]_i - \mathbb{E}_\mathcal{D} [r]_i \right) 
-  \left(\mathbb{E}_\mathcal{A} [\hat{r}]_i - \mathbb{E}_\mathcal{A} [r]_i \right) \right)  \nonumber  \\
& \quad \cdot  \left( \nabla_{\bm{x}_z} \mathbb{E}_\mathcal{D} [\hat{r}]_i - \nabla_{\bm{x}_z} \mathbb{E}_\mathcal{A} [\hat{r}]_i \right) \nonumber  \\
&=  \sum\nolimits_{i \in \mathcal{I}} \text{sign}   \left( \left(\mathbb{E}_\mathcal{D} [\hat{r}]_i - \mathbb{E}_\mathcal{D} [r]_i \right) 
-  \left(\mathbb{E}_\mathcal{A} [\hat{r}]_i - \mathbb{E}_\mathcal{A} [r]_i \right) \right)  \nonumber \\
\label{eqn_lastline} & \quad \cdot  \left( \frac{1}{\left| \mathcal{D}_i \right| } {\sum\nolimits_{u \in \mathcal{D}_i} \nabla_{\bm{x}_z} \hat{r}_{ui}}    -  \frac{1}{\left| \mathcal{A}_i \right| }{\sum\nolimits_{u \in \mathcal{A}_i} \nabla_{\bm{x}_z} \hat{r}_{ui}} \right),
\end{align}
where $(a)$ follows from the fact that $\mathbb{E}_\mathcal{D} [r]_i$ and  $\mathbb{E}_\mathcal{A} [r]_i$ are the average true rating scores for the item $i$ from disadvantaged and advantaged users, respectively, and do not change with the injected antidote users.

Then the key challenge boils down to how to compute $\nabla_{\bm{x}_z} \hat{r}_{ui}$ for $u \in \mathcal{U}, i \in \mathcal{I}$ in (\ref{eqn_lastline}).
Inspired by~\cite{li2016data,fang2020influence}, we let $\bm{H}_{\bm{x}_z}(\bm{p}_u)$ and $\bm{H}_{\bm{x}_z}(\bm{q}_i)$ denote the Jacobian matrices of $\bm{p}_u$ and $\bm{q}_i$ computed at $\bm{x}_z$, respectively. 
Since $\hat{r}_{ui} = \bm{p}_u^\top{\bm{q}_i}$, we have:
\begin{align}
\frac{{\partial {\hat{r}_{ui}}}}{{\partial {\bm{x}_z}}} = \bm{H}_{\bm{x}_z}(\bm{p}_u)^{\top} \bm{q}_i  + \bm{H}_{\bm{x}_z}(\bm{q}_i)^{\top} \bm{p}_u.
\end{align}

We next show how to exploit the first-order stationary condition to approximate $\bm{H}_{\bm{x}_z}(\bm{p}_u)$ and $\bm{H}_{\bm{x}_z}(\bm{q}_i)$.
Let $\bm{k} \in {\mathbb{R}^d}$ be the latent user vector for antidote user $z$.
When taking the current injected antidote user $z$ into consideration, we can rewrite the optimization problem of Eq.~(\ref{orig_matrix_opti_problem}) as  follows:
\begin{multline}
\label{mali_matrix_fact}
\mathop {\arg \min }\limits_{{\bm{P}},{\bm{Q}},\bm{k}} {\sum\nolimits_{(u,i) \in \Omega} {\left( {{r_{ui}} - \bm{p}_u^\top{\bm{q}_i}} \right)} ^2} 
+ {\sum\nolimits_{i \in \mathcal{I}} {\left( {{x_{zi}} - {\bm{k}^\top}{\bm{q}_i}} \right)} ^2} \\  
+ \lambda \left( {\sum\nolimits_{u \in \mathcal{U}} {{\lVert \bm{p}_u \rVert}_2^2 } } + {\lVert \bm{k} \rVert}_2^2+  {\sum\nolimits_{i \in \mathcal{I}} {{\lVert \bm{q}_i \rVert}_2^2 } }  \right),
\end{multline}
where $x_{zi}$ is the relaxed rating score that user $z$ gives to item $i$. 
Following~\cite{li2016data,fang2020influence}, the optimal solution of Problem (\ref{mali_matrix_fact}) satisfies the following conditions:
\begin{align} 
{\lambda}{\bm{p}_u} &= \sum\nolimits_{i \in \mathcal{I}_u} {({r_{ui}} - \bm{p}_u^\top{\bm{q}_i})} {\bm{q}_i} \label{KKT_p_u},
\\
{\lambda}{\bm{q}_i} &= \sum\nolimits_{u \in \mathcal{U}_i} {({r_{ui}} - \bm{p}_u^\top{\bm{q}_i})} {\bm{p}_u} + ({x_{zi}} - {\bm{k}^\top}{\bm{q}_i})\bm{k} \label{KKT_q_v},
\end{align}
where $\mathcal{I}_u$ is the set of items rated by user $u$, $\mathcal{U}_i$ is the set of users who rate the item $i$.
We further set the derivatives of (\ref{KKT_p_u})--(\ref{KKT_q_v}) with respect to $\bm{x}_z$ to zero.
Based on the fact $({\bm{p}_u^\top}\bm{q}_i)\bm{p}_u = (\bm{p}_u{\bm{p}_u^\top})\bm{q}_i$, we have $\frac{{\partial {\bm{p}_u}}}{{\partial {x_{zi}}}} = \mathbf{0}$ and:
\begin{align} 
\frac{{\partial {\bm{q}_i}}}{{\partial x_{zi}}} = {\left( \sum\nolimits_{u \in \mathcal{U}_i}{{\bm{p}_u}{\bm{p}_u^\top} + \bm{k}\bm{k}^\top} + {{\lambda}{\bm{I}}} \right)^{ - 1}}\bm{k},
\label{q_to_wzv}
\end{align}
where $\bm{I} \in {{\mathbb{R}}^{ d  \times d}}$ is the identity matrix. 
Further, we can obtain $\bm{H}_{\bm{x}_z}(\bm{p}_u)$ and $\bm{H}_{\bm{x}_z}(\bm{q}_i)$ by computing all items $i \in \mathcal{I}$ to finally obtain $\bm{g}_{\text{val}} (\bm{x}_z) $.
Also, we note that the gradients of other unfairness metrics (e.g., absolute, overestimation and non-parity unfairness metrics) can also be computed in a similar procedure.
To avoid repetition, the details about the derivation of gradients of other unfairness metrics are relegated to Appendix~\ref{appendix_gradient}.

After obtaining the gradient of loss function, we then use the projected gradient descent approach to determine the rating score vector $\bm{x}_{z}$ for antidote user $z$.
With $\bm{x}_{z}$, we select $n$ items with the largest absolute rating scores as filler items and truncate ratings to the set $\left\{r_{\text{min}}, \cdots, r_{\text{max}} \right\}$ (rounding rating score to the nearest integer value in the set $\left\{ r_{\text{min}}, \cdots, r_{\text{max}} \right\}$).
Note that we do not need to assign identities (e.g., gender, race, or age) to antidote users since the unfairness score is computed with original users with identities.
We summarize our antidote data generation model in Algorithm~\ref{attack_algo}.

\begin{algorithm}[t!]
	\caption{Our Antidote Data Generation Model.}\label{attack_algo}
	\begin{algorithmic}[1]
		\renewcommand{\algorithmicrequire}{\textbf{Input:}}
		\renewcommand{\algorithmicensure}{\textbf{Output:}}
		\REQUIRE User-item rating matrix $\bm{R}$, parameters $d, n, \alpha, \lambda$.
		\ENSURE  Antidote users set $ \mathcal{\widetilde{U}}$ and $\left\{ {{\bm{r}_z}} \right\}_{z \in \mathcal{\widetilde{U}}}$.
		\STATE Let $\mathcal{\widetilde{U}} \leftarrow \emptyset$. 
		
    	 //Add antidote users one by one.
		\FOR {each antidote user $z$} 
		\STATE Solve the optimization problem in Eq. (\ref{appproblem}) to get $\bm{x}_{z}$.
		\STATE Select $n$ items with the largest absolute values of $x_{zi}$ as filler items, and truncate ratings to the set $\left\{ 0, 1, \cdots, r_{\text{max}} \right\}$.
		\STATE Let $\mathcal{\widetilde{U}} \leftarrow \mathcal{\widetilde{U}} \cup \{ z \}$ and $R\leftarrow R \cup \{\bm{r}_z\}$.
		\ENDFOR
		\RETURN $\mathcal{\widetilde{U}}$ and $\left\{ {{\bm{r}_z}} \right\}_{z \in \mathcal{\widetilde{U}}}$.
	\end{algorithmic} 
\end{algorithm}

\subsection{FairRoad: Antidote Data Generation for Multi-Unfairness Optimization}

As mentioned in Section~\ref{sec:intro}, one may often prefer to simultaneously achieve good fairness performance in terms of different unfairness metrics.
In this section, we will show how to reduce the bias of multiple unfairness metrics.
For simplicity, suppose $M_1$ and $M_2$ are two unfairness metrics that the service provider wants to reduce (the same idea can be generalized to more than two fairness objectives).
A natural way to optimize $M_1$ and $M_2$ multi-objective optimization is to solve the following weighted-sum optimization problem:
\begin{align}
\label{multi_task_natural}
\begin{split}
& \min_{\bm{x}_{z}} \mathcal{L}(\bm{x}_{z}) = a \cdot M_1 + b\cdot  M_2 \\
& \text{   s.t. } x_{zi} \in [r_{\text{min}}, r_{\text{max}}].
\end{split}
\end{align}
where $a>0$ and $b>0$ are some fixed weight parameters that signify the relative importance/preference of the sub-objectives in performing multi-objective optimization.
However, solving the problem in Eq.~(\ref{multi_task_natural}) remains challenging in practice. 
Although it appears natural that Problem~(\ref{multi_task_natural}) can be solved by simply using gradient-descent-type approach, doing so is often inefficient.
This is due to the fact that many unfairness-related objective functions (cf. Eqs.~(\ref{Value_unfairness})--(\ref{parity_unfairness})) are often non-smooth and could have large local curvature.
As shown in~\cite{yu2020gradient,wang2020gradient}, for multi-objective optimization problems with conflicting sub-objective gradients (i.e., the cosine similarity between sub-objective gradients is negative), the gradient descent search process may easily stall and get stuck at ``local valley.''
To address this challenges, in this paper, we will propose a new approach inspired by~\cite{yu2020gradient,wang2020gradient} to effectively deflect the sub-objective gradients to reduce the unfairness for multiple unfairness metrics.

\begin{figure}[!t]
	\centering
	{\includegraphics[width= 0.4\textwidth]{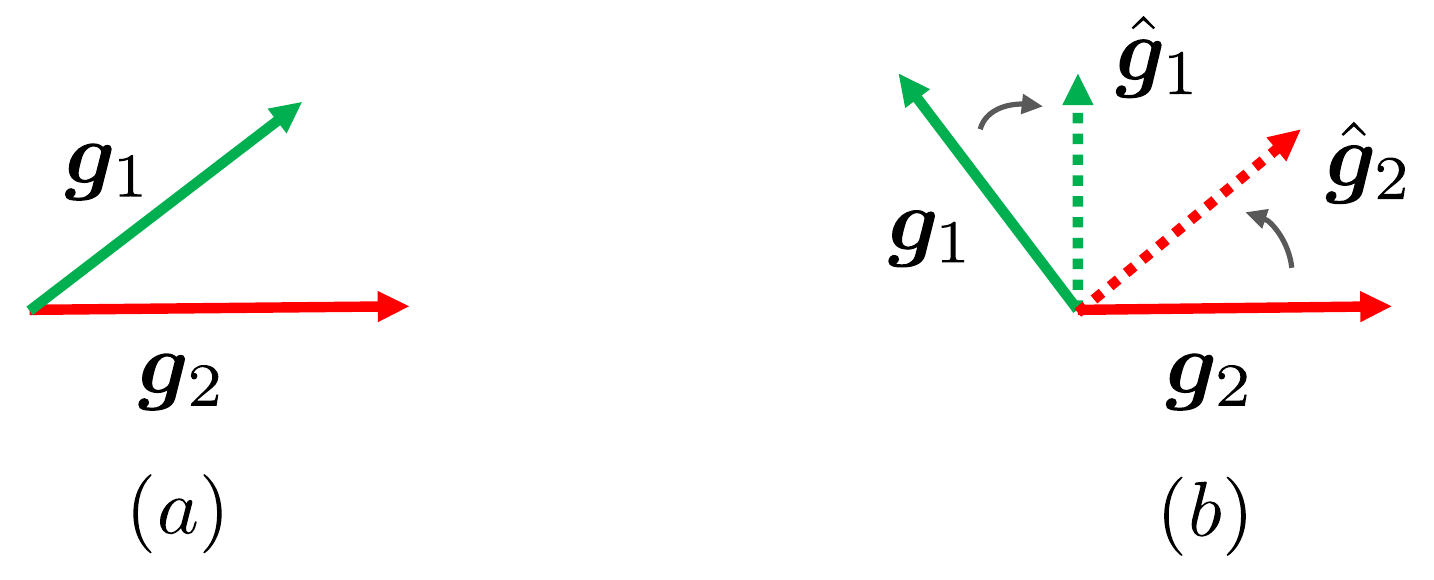}}
	\caption{Non-conflicting and conflicting gradients.}
	\label{confict_gradients}
\end{figure}

Specifically, we let $\bm{g}_1$ and $\bm{g}_2$ be the gradients of the unfairness sub-objective metrics $M_1$ and $M_2$ in some training iteration, respectively.
We first compute the cosine similarity between $\bm{g}_1$ and $\bm{g}_2$. 
If the cosine similarity is positive, as shown in Fig.~\ref{confict_gradients}(a), then gradients $\bm{g}_1$ and $\bm{g}_2$ are not in conflict, which implies that moving along the combined direction of $\bm{g}_1$ and $\bm{g}_2$ is unlikely being trapped, particularly at points with a large local curvature.
In this case, one can simply use the gradient of the weighted-sum objective function:
\begin{align}
\label{eq_no_conflict_grad}
\hat{\bm{g}} =  a \bm{g}_1 + b \bm{g}_2.
\end{align}

On the other hand, if the cosine similarity is negative, i.e., $\bm{g}_1$ and $\bm{g}_2$ are pointing toward opposite directions over the local manifold of the combined objective function.
To avoid being trapped at points with large local curvature, we alter $\bm{g}_1$ by projecting $\bm{g}_1$ onto the normal plane of $\bm{g}_2$, see the green dashed line in Fig.~\ref{confict_gradients}(b).
The projected gradient $\hat{\bm{g}}_1$ can be computed as:
\begin{align}
\hat{\bm{g}}_1 = \bm{g}_1 - \frac{\left\langle {\bm{g}_1, \bm{g}_2} \right\rangle}{\left\| \bm{g}_2 \right\|_2^2} \bm{g}_2.
\end{align}
Similarly, we replace $\bm{g}_2$ by its projection onto the normal plane of $\bm{g}_1$, see the red dashed line in Fig.~\ref{confict_gradients}(b).
Similarly, the projected gradient $\hat{\bm{g}}_2$ can be computed as:
\begin{align}
\hat{\bm{g}}_2 = \bm{g}_2 - \frac{\left\langle {\bm{g}_1, \bm{g}_2} \right\rangle}{\left\| \bm{g}_1 \right\|_2^2} \bm{g}_1.
\end{align}
The final combined gradient can be calculated as the weighted sum of two altered gradients:
\begin{align}
\label{eq_conflict_grad}
\hat{\bm{g}} 
= a \hat{\bm{g}}_1 + b \hat{\bm{g}}_2 
= a \bm{g}_1  + b \bm{g}_2 - a \frac{\left\langle {\bm{g}_1, \bm{g}_2} \right\rangle}{\left\| \bm{g}_2 \right\|_2^2} \bm{g}_2 - b \frac{\left\langle {\bm{g}_1, \bm{g}_2} \right\rangle}{\left\| \bm{g}_1 \right\|_2^2} \bm{g}_1.
\end{align}

To summarize, given two unfairness metrics that the service provider wants to minimize, in each training step, we first compute the gradients of the sub-objectives corresponding to the two unfairness metrics, and then compute the final gradient $\hat{\bm{g}} $ according to Eq.~(\ref{eq_no_conflict_grad}) or Eq.~(\ref{eq_conflict_grad}).
With $\hat{\bm{g}} $, we can again use the projected gradient approach to update rating score for each antidote user. 
Note that our proposed FairRoad framework could be easily extended to multi-unfairness metrics (more than two unfairness metrics). 
The way of how to deflect more than two gradients could be found in~\cite{yu2020gradient,wang2020gradient}, and we omit it here for briefly.

\section{Experiments} \label{sec:exp}

\begin{table}[t!]
	\centering
	\caption{Rating and observation block models for synthetic dataset.}
	\small
	\label{syn_gen_block}
	\subfloat[Rating block model.]
	{
		\begin{tabular}{|l|c|c|}
			\toprule
			& \multicolumn{1}{l|}{STEM} & \multicolumn{1}{l|}{Non-STEM} \\
			\hline
			Male  & $\alpha_1$     & $\alpha_2$ \\
			\hline
			Female & $\alpha_3$     & $\alpha_4$ \\
			\bottomrule
		\end{tabular}%
	}
	\quad
	\subfloat[Observation block model.]
	{
		\begin{tabular}{|l|c|c|}
			\toprule
			& \multicolumn{1}{l|}{STEM} & \multicolumn{1}{l|}{Non-STEM} \\
			\hline
			Male  & $\beta_1$     & $\beta_2$ \\
			\hline
			Female & $\beta_3$     & $\beta_4$ \\
			\bottomrule
		\end{tabular}%
	}
\end{table}%

\begin{table*}[t!]
	\centering
	\addtolength{\tabcolsep}{-0.55pt}
	\setlength{\doublerulesep}{3\arrayrulewidth}
	\caption{Unfairness scores under different methods on synthetic and Yelp datasets. 
    The synthetic dataset exhibits no population imbalance ($\alpha_1=\alpha_2=0.4$) and has observation bias ($\beta_1=0.2$, $\beta_2=0.1$).
	}
	\label{results_attack_size}
	\footnotesize
	\begin{tabular}{l|l|cccc|cccc|cccc|cccc}
		\toprule
	    \multirow{3}[6]{*}{Datasets} & \multirow{3}[6]{*}{Method} & \multicolumn{16}{c}{Fraction of antidote users} 
	    \bigstrut\\
		\cline{3-18}           &
		&  \multicolumn{4}{c|}{Value unfairness}      
		& \multicolumn{4}{c|}{Absolute unfairness}        
		&  \multicolumn{4}{c|} {Overestimation unfairness}       
		 & \multicolumn{4}{c}{Non-parity unfairness} 
		\bigstrut \\
		\cline{3-18}          &       & 0.5\%     & 1\%     & 2\%    & 3\%    & 0.5\%     & 1\%     & 2\%    & 3\%    & 0.5\%     & 1\%     & 2\%    & 3\%    & 0.5\%     & 1\%       & 2\%    & 3\% 
		\bigstrut \\
		\hline
		\hline
		\multirow{7}[2]{*}{Synthetic} 
		& None &0.280 & 0.280 &0.280 & 0.280 & 0.107 & 0.107 & 0.107 & 0.107 & 0.144 & 0.144 & 0.144 & 0.144 &0.041 & 0.041 & 0.041 &0.041 \\
		& Regularization &  0.252    &  0.252 &  0.252  &  0.252   &  0.104    & 0.104  & 0.104   &  0.104  & 0.127  & 0.127  & 0.127  & 0.127  & 0.040  & 0.040   &  0.040   & 0.040\\
		& Maximum & 0.281 & 0.290 &0.259 & 0.285 &0.104 &0.106  & 0.108  & 0.117    &  0.135  & 0.132   &  0.128   & 0.126 & 0.042  & 0.054  & 0.062  & 0.056  \\
		& Minimum & 0.292 &0.282 & 0.256  &0.272 & 0.105  &  0.105  &  0.106   &  0.124 &  0.136   & 0.128 & 0.124  & 0.124 & 0.055 & 0.044 &  0.064  & 0.049 \\
		& Random & 0.275  & 0.276 &  0.288 &0.263 &  0.105  & 0.104  & 0.116  &  0.108  &  0.142 & 0.144  &  0.149 & 0.139 & 0.044 &  0.044  & 0.043   &  0.042 \\
		& Rastegarpanah et al. & 0.252  & 0.244 &  0.226 & 0.211 &  0.104  & 0.104  & 0.101  &  0.096  &  0.136 & 0.134  &  0.116 & 0.116 & 0.052 &  0.041  &0.040   &  0.038 \\
		& FairRoad & \textbf{0.236} & \textbf{0.209} & \textbf{0.199} & \textbf{0.158} & \textbf{0.101}  & \textbf{0.097} & \textbf{0.086} & \textbf{0.082}  & \textbf{0.124} & \textbf{0.124} & \textbf{0.099} & \textbf{0.091} & \textbf{0.039} & \textbf{0.037} &  \textbf{0.032} & \textbf{0.032}\\
		\hline
		\multirow{7}[2]{*}{Yelp} 
		& None & 0.126 & 0.126 & 0.126 & 0.126 & 0.030 & 0.030 & 0.030 & 0.030  & 0.053 & 0.053 & 0.053 & 0.053 & 0.033 & 0.033 & 0.033 & 0.033 \\
		& Regularization & 0.123 &  0.123 &  0.123 &  0.123  &  \textbf{0.028}  &  0.028  &  0.028  &  0.028&  0.051  &  0.051&0.051  &  0.051& 0.029   &  0.029 & 0.029  & 0.029 \\
		& Maximum & 0.129  & 0.127 & 0.130 & 0.129 & 0.036 & 0.038  & 0.042  &  0.042  & 0.053  &  0.055 & 0.057   & 0.056 & 0.033 & 0.033  & 0.033 & 0.035 \\
		& Minimum & 0.128 & 0.125 & 0.125 & 0.124 &0.032  & 0.030 &  0.030 &  0.029   & 0.053 & 0.053 & 0.055  & 0.056 & 0.033 &0.031  & 0.033   &0.033 \\
		& Random & 0.132 & 0.130 & 0.130 &0.130 &  0.035  & 0.038 & 0.041 &  0.046 & 0.053   & 0.055  &  0.058  &  0.056  &  0.030  & 0.032  & 0.032 & 0.033\\
		& Rastegarpanah et al. & 0.125  & 0.121 &  0.117 & 0.116 &  0.030  & 0.029  & 0.029  &  0.028  &  0.052 & 0.051  &  0.051 & 0.050 & 0.030 &  0.030  & 0.028   &  0.027 \\
		& FairRoad & \textbf{0.119} & \textbf{0.115} & \textbf{0.102} & \textbf{0.099} & \textbf{0.028} & \textbf{0.027} & \textbf{0.027} & \textbf{0.026} & \textbf{0.049} & \textbf{0.049} & \textbf{0.048} & \textbf{0.047} & \textbf{0.028} & \textbf{0.025} & \textbf{0.025} & \textbf{0.025} \\
		\bottomrule
	\end{tabular}%
\end{table*}%

\subsection{Experimental Settings}

{\bf 1) Datasets:}
In our paper, we use synthetic and the Yelp datasets to show the effectiveness of our FairRoad approach.
For the synthetic data, we construct two user groups $\left\{  \text{Male}, \text{Female}  \right\}$ and two item groups $\left\{  \text{STEM}, \text{Non-STEM}  \right\}$, where $\text{STEM}$ represents the science, technology, engineering, and mathematics courses, and $\text{Non-STEM}$ denotes the Non-STEM courses.
Inspired by~\cite{yao2017beyond,cho2021equal}, we consider the \textit{population imbalance} and \textit{observation bias} when generating the synthetic dataset.
Population imbalance occurs when different types of users have different preferences to give high ratings to different types of items.
For instance, male students may be more likely to give high ratings to STEM courses compared with female students.
Observation bias happens when different types of users have different tendencies to rate different types of items, e.g., male students rate STEM courses with higher frequency relative to females.

We generate the synthetic dataset with binary ratings, i.e, $r_{ui} \in \left\{ -1, 1 \right\}$, where $-1$ and $1$ denote user $u$ dislikes and likes item $i$, respectively.
To simulate the population imbalance, we generate a rating matrix according to the rating block model as shown in Table~\ref{syn_gen_block}(a).
In Table~\ref{syn_gen_block}(a), each value is a probability that a user in a user group likes an item in an item group.
For instance, a male student likes a STEM course with probability $\alpha_1$ and  dislikes a STEM course with probability 1-$\alpha_1$. 
We use a observation block model to simulate the observation bias, as shown in Table~\ref{syn_gen_block}(b), where each value is a probability that a user in a user group gives rating score (-1 or 1) to an item in this item group.
For instance, a male student rates a STEM course with probability $\beta_1$.
To simplify the simulation, we let $\alpha_1=\alpha_4$, $\alpha_2=\alpha_3$, $\beta_1=\beta_4$ and $\beta_2=\beta_3$.
Thus, we can use $\alpha_1$ and $\alpha_2$ parameters to control the population imbalance, and $\beta_1$ and $\beta_2$ to determine the observation bias. Note that when $\alpha_1= \alpha_2$, the dataset exhibits no population imbalance, and the dataset exhibits no observation bias if $\beta_1= \beta_2$.
For synthetic data, we randomly generate 800 users, which include 400 males and 400 females; we also randomly generate 600 items, among which 300 items belong to STEM and 300 items belong Non-STEM. By default, the synthetic dataset exhibits no population imbalance ($\alpha_1=\alpha_2=0.4$), and has observation bias ($\beta_1=0.2$, $\beta_2=0.1$). The total number of rating data is 71,630. 
The Yelp~\cite{yelpURL} dataset contains 6,931 users, 5,494 items and 175,105 ratings ranging from 1 to 5. We use gender api~\cite{gender_api} tool to extract identities with 574 males and 742 females.
We treat female users as advantaged users, and male users as disadvantaged users.

\smallskip
\noindent
{\bf 2) Baseline Models:} 
To demonstrate the effectiveness of FairRoad, we compare FairRoad with the following baseline methods. 
For the five baselines, Regularization approach does not inject extra data into the systems, while the other four baselines (Maximum, Minimum, Random, Rastegarpanah et al.~\cite{rastegarpanah2019fighting}) improve fairness by injecting antidote data into the systems.

\myparatight{2-a) Regularization~\cite{yao2017beyond}} This method improves fairness by adding a regularization term to the learning objective (treating an unfairness metric as a regularization term).

\myparatight{2-b) Maximum data generation (Maximum)} In this antidote data generation method, the service provider randomly selects some items as fillers for each antidote user, and always provides the maximum rating for each filler item.

\myparatight{2-c) Minimum data generation (Minimum)} 
The service provider randomly selects some items as filler items for each antidote user, and always provides the minimum rating for each filler item.

\myparatight{2-d) Random data generation (Random)} The service provider selects filler items for each antidote user uniformly at random, and assigns rating from the set $\left\{ r_{\text{min}}, \cdots, r_{\text{max}} \right\}$ to each filler item uniformly at random.

\myparatight{2-e) Rastegarpanah et al.~\cite{rastegarpanah2019fighting}} The authors in~\cite{rastegarpanah2019fighting} propose one optimization based approach to improve the polarization and fairness of recommender systems   by adding more data to the input.
The service provider in~\cite{rastegarpanah2019fighting} injects all antidote users at one time. Specifically, the service provider first selects $n$ filler items uniformly at random for each antidote user to rate, then generates rating scores for the filler items by solving an optimization problem.

\smallskip
\noindent
{\bf 3) Unfairness Metrics:} 
In our experiments, we use value unfairness, absolute unfairness, overestimation unfairness and non-parity unfairness as our unfairness metrics.
In order to measure the effectiveness of our FairRoad method, we use the unfairness score as our evaluation metric.
The unfairness score is computed as Eq.~(\ref{Value_unfairness}), Eq.~(\ref{Absolute_unfairness}), Eq.~(\ref{Over_unfairness}) or Eq.~(\ref{parity_unfairness}) according to different unfairness metrics.
The smaller the score, the fairer the recommender systems.

\smallskip
\noindent
{\bf 4) Parameter Setting:} 
By default, we use the following default parameters: the size of antidote users is 2\% of the original users.
The number of filler items is $n=200$.
We set $d=8, \lambda=0.1, a=1,b=1$.

\begin{figure*}[!t]
	\centering
	\includegraphics[scale = 0.51]{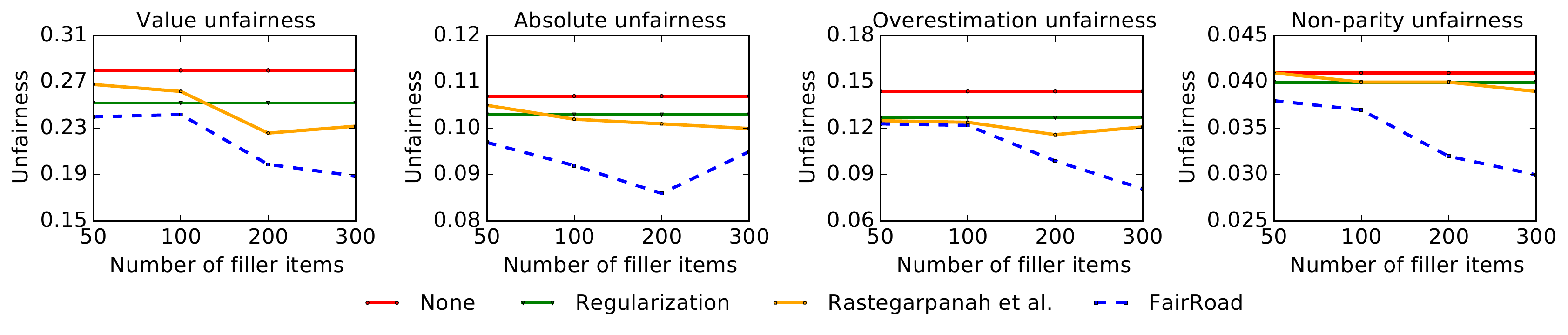}
	\caption{Impact of the number of filler items on the synthetic dataset.
	}
	\label{synthetic_filler_size}
\end{figure*}

\begin{figure*}[!t]
	\centering
	\includegraphics[scale = 0.51]{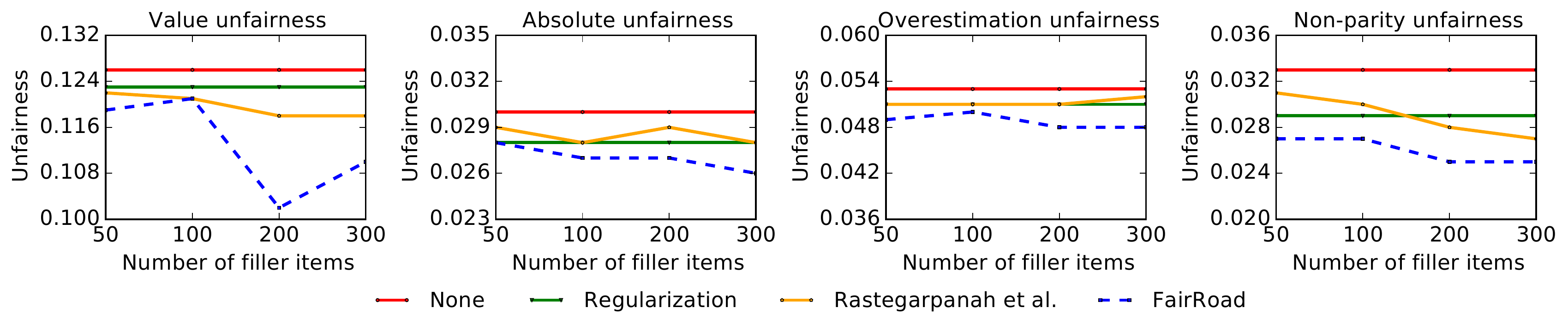}
	\caption{Impact of the number of filler items on the Yelp dataset.}
	\label{yelp_filler_size}
\end{figure*}

 \begin{figure*}[!t]
 	\centering
 	\includegraphics[scale = 0.51]{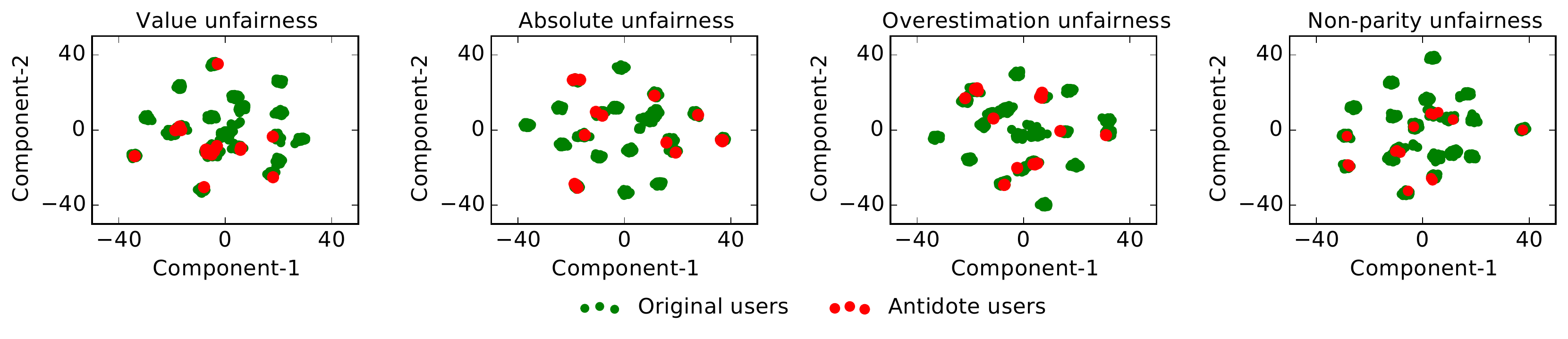}
 	\caption{Original and antidote users in the representation space on the synthetic dataset. 
 	}
 	\label{synthetic_tsne}
 \end{figure*}

\subsection{Experimental Results}

\myparatight{1) Impact of antidote users size}
Table~\ref{results_attack_size} shows the unfairness scores of different unfairness metrics under different approaches.
The synthetic dataset (default synthetic dataset) in Table~\ref{results_attack_size} exhibits no population imbalance ($\alpha_1=\alpha_2=0.4$), and has observation bias ($\beta_1=0.2$, $\beta_2=0.1$).
In Table~\ref{results_attack_size}, 
``None'' means the unfairness score computed with the original rating data and does not incorporate the unfairness metric as a regularization term.
Additional experimental results for different settings of synthetic datasets are relegated to Tables~\ref{results_Synthetic_two}-\ref{results_Synthetic_five} in Appendix~\ref{appendix_Synthetic}.
Table~\ref{results_Synthetic_two} summaries the results when the synthetic dataset exhibits no population imbalance but different observation bias.
Table~\ref{results_Synthetic_three} shows the results when the synthetic dataset exhibits different population imbalance and has no observation bias.
Table~\ref{results_Synthetic_five} illustrates the results when the synthetic dataset exhibits population imbalance and also has observation bias.

From Table~\ref{results_attack_size}, we observe that our proposed FairRoad approach effectively reduces the unfairness of recommender systems with only a small amount of antidote data.
For instance, FairRoad reduces 31.25\% (the unfairness score decreases from 0.144 to 0.099) of overestimation unfairness when injecting only 2\% of antidote users into the synthetic dataset.
Also, a 19.05\% reduction of value unfairness is reduced with only 2\% of antidote users on Yelp dataset (the unfairness score decreases from 0.126 to 0.102).
Second, FairRoad outperforms baseline methods for four unfairness metrics in most cases.
Maximum, Minimum and Random data generation approaches could not reduce the unfairness or may even increase unfairness.
The Regularization method only slightly reduces the unfairness of recommender systems.
The approach proposed in Rastegarpanah et al.~\cite{rastegarpanah2019fighting} only achieves sub-optimal performance since it randomly selects $n$ items as filler items for each antidote user.
We also note that FairRoad improves the fairness without sacrificing the performance.
For instance, on synthetic data, the Root Mean Square Error (RMSE) without antidote data is 0.630, and the RMSE after injecting 2\% of antidote data generated by FairRoad based on value, absolute, overestimation and non-parity unfairness metrics are 0.633, 0.631, 0.630 and 0.630, respectively. 
On Yelp dataset, the RMSE without antidote data is 0.697, the RMSE are 0.698, 0.701, 0.697 and 0.698 with 2\% of antidote data generated based on four unfairness notions.

\myparatight{2) Impact of the number of filler items} 
Figures~\ref{synthetic_filler_size}-\ref{yelp_filler_size} illustrate the results of FairRoad with different number of filler items on the synthetic and Yelp datasets, where the fraction of antidote users is set to 2\%.
We do not investigate the impact of number of filler items on Maximum, Minimum and Random data generation approaches since these three methods almost could not reduce the unfairness.
For the synthetic dataset, the unfairness score decreases with more filler items for the overestimation and non-parity unfairness metrics.
However, the unfairness score does not always decrease with more filler items, implying that there exists an optimal amount of filler items.
For instance, on the Yelp dataset, FairRoad performs the best when the number of filler items is 200.
The unfairness score then increases as more filler items are used.
We also observe that FairRoad outperforms the Regularization and Rastegarpanah et al.~\cite{rastegarpanah2019fighting} methods in most test cases.

\myparatight{3) Antidote users in the representation space}
Even though the service provider itself injects some antidote data into the recommender systems to achieve fairness, recent studies have shown that recommender systems are vulnerable to poisoning attack~\cite{fang2018poisoning,li2016data}, where the attacker could inject some fake users into the systems to degrade the performance of recommender systems.
The service provider may arm the recommender systems with anomalies detection capability to detect the fake users and remove the detected fake users from the systems.
Based on this observation, the antidote users generated by FairRoad should mimic the rating behavior of original users such that the anomaly detection algorithm deployed by the system will not consider the antidote users as anomalies. 
In this part, we check whether the antidote users generated by FairRoad can be detected as anomalies in the representation space.
Figure~\ref{synthetic_tsne} illustrates the original users and antidote users generated by FairRoad in the representation space for four unfairness metrics on synthetic dataset, where the fraction of antidote users is set to 2\% and the plots are visualized by the t-SNE~\cite{van2008visualizing} technique.
We see that the antidote users generated by FairRoad distributed evenly in the representation space, and it is hard to distinguish the antidote users from original users via distribution discrepancy.

\myparatight{4) Transferability of FairRoad}
In this experiment, we study the transferability of FairRoad, i.e., generating antidote data based on one unfairness metric could also reduce the unfairness of other unfairness metrics.
Here, the fraction of antidote users is set to 2\% and the results are shown in Table~\ref{results_transferability}.
In Table~\ref{results_transferability}, each row is the source unfairness metric, based on which we generate the antidote data, each column is the targeted unfairness metric.
For instance, we generate antidote data according to the value unfairness metric, then compute the unfairness score of absolute unfairness notion after injecting those antidote data into the systems.
``Value'' in Table~\ref{results_transferability} means the value unfairness metric.
We observe that FairRoad has good transferability in most cases.
For instance, on the synthetic dataset, the unfairness score of absolute unfairness metric decreases from 0.107 without antidote data (see Table~\ref{results_attack_size}) to 0.099 with 2\% of antidote users generated by value unfairness notion.

\myparatight{5) Multiple unfairness metrics optimization}
In this experiment, we evaluate the performance of FairRoad in optimizing two unfairness metrics simultaneously.
Here, the fraction of antidote users is set to 2\% and the results are shown in Table~\ref{results_multiple}, where ``Value+Absolute'' means FairRoad optimizes value and absolute unfairness metrics simultaneously.
We observe that FairRoad reduces the unfairness of multiple unfairness metrics at the same time.
For instance, on the synthetic dataset, when FairRoad optimizes value and overestimate unfairness metrics simultaneously.
The unfairness score of value unfairness decreases from 0.280 without antidote data (see Table~\ref{results_attack_size}) to 0.204 with 2\% of antidote users; the unfairness score of overestimate unfairness decreases from 0.144 without antidote data (see Table~\ref{results_attack_size}) to 0.104 with only 2\% of antidote users.

To see the benefits of deflecting conflicting sub-objective gradients in our FairRoad approach, we also consider the setting where one optimizes two unfairness metrics simultaneously without resolving conflicting gradients, i.e., computes the final gradient only based on Eq.~(\ref{eq_no_conflict_grad}).
The results of optimizing non-parity and other unfairness metrics simultaneously on the synthetic data are shown in Table~\ref{results_multiple_no_deconflict}, where the fraction of antidote users is set to 2\%.
We observe that, numerically, this approach could increase the unfairness score of targeted unfairness metric.
For instance, on the synthetic dataset, the unfairness score of overestimation unfairness decreases from 0.144 without antidote data to 0.104 and with 2\% of antidote users.
However, the unfairness score of non-parity increases from 0.041 (no antidote data) to 0.052 (2\% of antidote users).

\begin{table}[!t]
	\centering
	\addtolength{\tabcolsep}{-0.7pt}
	\setlength{\doublerulesep}{3\arrayrulewidth}
	\caption{Transferability of FairRoad.}
	\label{results_transferability}
	\footnotesize
	\begin{tabular}{l|l|c|c|c|c}
		\toprule
		\multirow{1}{*}{Datasets} & \multirow{1}{*}{Source metric} 
		&  \multicolumn{1}{c|}{Value}        & \multicolumn{1}{c|}{Absolute}        &  \multicolumn{1}{c|}{Overestimation }        & \multicolumn{1}{c}{Non-parity} 
		\bigstrut \\
		\hline
		\hline
		\multirow{4}[2]{*}{Synthetic} 
		& Value &  \textbf{0.199} & 0.099  & 0.123 & 0.046  \\
		& Absolute   & 0.275  & \textbf{0.086}  &  0.141     & 0.045     \\
		& Overestimation &  0.246  & 0.105   &  \textbf{0.099}  &  0.057  \\
		& Non-parity & 0.285   & 0.107  &   0.134 &  \textbf{ 0.032} \\
		\hline
		\multirow{4}[2]{*}{Yelp} 
		& Value & \textbf{0.102}  &  0.028 & 0.051 & 0.033 \\
		& Absolute   &  0.122  &  \textbf{0.027}     &  0.051     &  0.033    \\
		& Overestimation & 0.121   &  0.028 & \textbf{0.048}   &  0.033  \\
		& Non-parity &  0.132  & 0.030  & 0.054   & \textbf{0.025}   \\
		\bottomrule
	\end{tabular}%
\end{table}%

\begin{table}[!t]
	\centering
	\addtolength{\tabcolsep}{-1.2pt}
	\setlength{\doublerulesep}{3\arrayrulewidth}
	\caption{Unfairness scores under FairRoad method on synthetic and Yelp datasets, where FairRoad optimizes two unfairness metrics simultaneously.}
	\label{results_multiple}
	\footnotesize
	\begin{tabular}{l|l|c|c|c|c}
		\toprule
		\multirow{1}{*}{Datasets} & \multirow{1}{*}{Source metric} 
		&  \multicolumn{1}{c|}{Value}       
	    & \multicolumn{1}{c|}{Absolute}      
	 	&  \multicolumn{1}{c|}{\thead{Over-\\ estimation} }        
	    & \multicolumn{1}{c}{\thead{Non-\\ parity}} 
		\bigstrut \\
		\hline
		\hline
		\multirow{6}[2]{*}{Synthetic} 
		& Value+Absolute & 0.208 & \textbf{0.093} & 0.121 & 0.050\\
		& Value+Overestimation   & \textbf{0.204}   &  0.097  & \textbf{0.104} &  0.067  \\
		& Value+Non-parity  & 0.224  &  0.105 &  0.109  & 0.037   \\
		& Absolute+Overestimation & 0.247  &  0.094 & 0.117 & 0.054   \\
		& Absolute+Non-parity & 0.263  &  0.097  &0.135  & 0.036 \\
		& Overestimation+Non-parity &  0.249 & 0.099 &0.108 & \textbf{0.035} \\
		\hline
		\multirow{6}[2]{*}{Yelp} 
		& Value+Absolute & \textbf{0.116}  &  0.028  &0.053 & 0.033  \\
		& Value+Overestimation   &   0.121   & 0.030 &   0.050    &   0.033   \\
		& Value+Non-parity  & 0.119   &  0.033 & 0.053   & \textbf{ 0.027}  \\
		& Absolute+Overestimation &  0.125  & 0.029  &  \textbf{0.048}  &  0.034  \\
		& Absolute+Non-parity & 0.122   &  \textbf{0.027}  &  0.051 & 0.031  \\
		& Overestimation+Non-parity &  0.122  & 0.030  & 0.050  &  0.031   \\
		\bottomrule
	\end{tabular}%
\end{table}%

\begin{table}[!t]
	\centering
	\addtolength{\tabcolsep}{-1.2pt}
	\setlength{\doublerulesep}{3\arrayrulewidth}
	\caption{Unfairness scores under FairRoad method on synthetic dataset, where FairRoad does not deflect gradients when optimizes two unfairness metrics.}
	\label{results_multiple_no_deconflict}
	\footnotesize
	\begin{tabular}{l|l|c|c|c|c}
		\toprule
		\multirow{1}{*}{Dataset} & \multirow{1}{*}{Source metric} 
		&  \multicolumn{1}{c|}{Value}       
		& \multicolumn{1}{c|}{Absolute}      
		&  \multicolumn{1}{c|}{\thead{Over-\\ estimation} }        
		& \multicolumn{1}{c}{\thead{Non-\\ parity}} 
		\bigstrut \\
		\hline
		\hline
		\multirow{3}[2]{*}{Synthetic} 
		& Value+Non-parity  & 0.236  &  0.108 &  0.141  & \textbf{0.036}   \\
		& Absolute+Non-parity & 0.264  &  0.112  &0.135  &  \textbf{0.036}  \\
		& Overestimation+Non-parity &  \textbf{0.233} & \textbf{0.102} & \textbf{0.104} & 0.052 \\
		\bottomrule
	\end{tabular}%
\end{table}


\section{Conclusion} \label{sec:conclusion}
In this paper, we proposed an effective method named FairRoad to achieve fairness for recommender systems.
In our FairRoad approach, the key idea is that the service provider can inject a small number of antidote data into the recommender systems to improve the fairness of a targeted recommender system.
We formulate the antidote data generation task as an optimization problem, the solution of which provides the ratings for the injected antidote users.
We further proposed a gradient-deflecting technique to reduce the unfairness of multiple unfairness metrics simultaneously.
Extensive experiments on four synthetic and the real-world Yelp datasets demonstrated that our FairRoad approach significantly improves the fairness of recommender systems with a small amounts of antidote data.

\section*{Acknowledgements}
This work has been supported in part by NSF grants CAREER CNS-2110259, CNS-2112471, CNS-2102233, CCF-2110252, and a Google Faculty Research Award.

\balance
\bibliographystyle{ACM-Reference-Format}
\bibliography{refs,Kevin_FAI}

\appendix

\appendix

\section{Derivation of the gradients of unfairness metrics} \label{appendix_gradient}

\subsection{Gradient of absolute unfairness metric}

\begin{align}
&\frac{1}{\left| \mathcal{I} \right|} \sum\nolimits_{i \in \mathcal{I}} \nabla_{\bm{x}_{z}} \left| \left|\mathbb{E}_\mathcal{D} [\hat{r}]_i - \mathbb{E}_\mathcal{D} [r]_i \right|
-  \left|\mathbb{E}_\mathcal{A} [\hat{r}]_i - \mathbb{E}_\mathcal{A} [r]_i \right| \right| \nonumber  \\
&= \frac{1}{\left| \mathcal{I} \right|} \sum\nolimits_{i \in \mathcal{I}} \text{sign}   \left( \left|\mathbb{E}_\mathcal{D} [\hat{r}]_i - \mathbb{E}_\mathcal{D} [r]_i \right| 
-  \left|\mathbb{E}_\mathcal{A} [\hat{r}]_i - \mathbb{E}_\mathcal{A} [r]_i \right| \right)  \nonumber  \\
& \quad \cdot \left( \nabla_{\bm{x}_{z}} \left|\mathbb{E}_\mathcal{D} [\hat{r}]_i - \mathbb{E}_\mathcal{D} [r]_i \right|  -  
\nabla_{\bm{x}_{z}} \left|\mathbb{E}_\mathcal{A} [\hat{r}]_i - \mathbb{E}_\mathcal{A} [r]_i \right|   \right)     \nonumber  \\
&= \frac{1}{\left| \mathcal{I} \right|} \sum\nolimits_{i \in \mathcal{I}} \text{sign}   \left( \left|\mathbb{E}_\mathcal{D} [\hat{r}]_i - \mathbb{E}_\mathcal{D} [r]_i \right| 
-  \left|\mathbb{E}_\mathcal{A} [\hat{r}]_i - \mathbb{E}_\mathcal{A} [r]_i \right| \right)  \nonumber  \\
& \quad \cdot \left( \text{sign} (\mathbb{E}_\mathcal{D} [\hat{r}]_i - \mathbb{E}_\mathcal{D} [r]_i ) \cdot \frac{1}{\left| \mathcal{D}_i \right| } {\sum\nolimits_{u \in \mathcal{D}_i} \nabla_{\bm{x}_{z}} \hat{r}_{ui}}    \right. \nonumber\\
& \quad\quad\quad \left. -  
\text{sign} (\mathbb{E}_\mathcal{A} [\hat{r}]_i - \mathbb{E}_\mathcal{A} [r]_i ) \cdot  \frac{1}{\left| \mathcal{A}_i \right| } {\sum\nolimits_{u \in \mathcal{A}_i} \nabla_{\bm{x}_{z}} \hat{r}_{ui}}    \right)     
\end{align}

\subsection{Gradient of overestimation unfairness metric}
Let $C_3 = \text{max} \left\{0,  \mathbb{E}_\mathcal{D} [\hat{r}]_i - \mathbb{E}_\mathcal{D} [r]_i \right\}$ and $C_4 = \text{max} \left\{0,  \mathbb{E}_\mathcal{A} [\hat{r}]_i - \mathbb{E}_\mathcal{A} [r]_i \right\}$,
then we have:
\begin{align}
& \frac{1}{\left| \mathcal{I} \right|} \sum\nolimits_{i \in \mathcal{I}} \nabla_{\bm{x}_{z}} \left| C_3 - C_4   \right|  \nonumber  \\
& = \frac{1}{\left| \mathcal{I} \right|} \sum\nolimits_{i \in \mathcal{I}} \text{sign}( C_3 - C_4) \cdot \left(  \nabla_{\bm{x}_{z}}C_3 -  \nabla_{\bm{x}_{z}}C_4  \right), 
\end{align}
where 
$\nabla_{\bm{x}_{z}}C_3$ and $\nabla_{\bm{x}_{z}}C_4$ can be computed as:
\begin{equation}
\nabla_{\bm{x}_{z}}C_3 =
\left\{
\begin{matrix}
\frac{1}{\left| \mathcal{D}_i \right| }{\sum\nolimits_{u \in \mathcal{D}_i} \nabla_{\bm{x}_{z}} \hat{r}_{ui}} , &  \mathbb{E}_\mathcal{D} [\hat{r}]_i - \mathbb{E}_\mathcal{D} [r]_i >0,\\
0, &  \text{otherwise}.
\end{matrix}
\right.
\end{equation}

\begin{equation}
\nabla_{\bm{x}_{z}}C_4 =
\left\{
\begin{matrix}
\frac{1}{\left| \mathcal{A}_i \right| }{\sum\nolimits_{u \in \mathcal{A}_i} \nabla_{\bm{x}_{z}} \hat{r}_{ui}} , &  \mathbb{E}_\mathcal{A} [\hat{r}]_i - \mathbb{E}_\mathcal{A} [r]_i >0,\\
0, &  \text{otherwise}.
\end{matrix}
\right.
\end{equation}

\subsection{Gradient of non-parity unfairness metric}
\begin{align}
\!\!\!\!\!  &\nabla_{\bm{x}_{z}} \left| \mathbb{E}_\mathcal{D} [\hat{r}] - \mathbb{E}_\mathcal{A} [\hat{r}]  \right|   \nonumber  \\
& = \text{sign}   \left( \mathbb{E}_\mathcal{D} [\hat{r}] - \mathbb{E}_\mathcal{A} [\hat{r}] \right) \cdot \left( \nabla_{\bm{x}_{z}}\mathbb{E}_\mathcal{D} [\hat{r}] -  \nabla_{\bm{x}_{z}}\mathbb{E}_\mathcal{A} [\hat{r}]   \right)  \nonumber  \\
& = \text{sign}   \left( \mathbb{E}_\mathcal{D} [\hat{r}] \!-\! \mathbb{E}_\mathcal{A} [\hat{r}] \right) \! \cdot \!  \left( \frac{1}{C_5} \!\! \sum\limits_{u \in \mathcal{D}}\sum\limits_{i \in \mathcal{I}_u} \nabla_{\bm{x}_{z}} \hat{r}_{ui} 
 \!-\!  \frac{1}{C_6} \!\! \sum\limits_{u \in \mathcal{A}}\sum\limits_{i \in \mathcal{I}_u} \nabla_{\bm{x}_{z}} \hat{r}_{ui}\right),  
\end{align}
where $C_5$ and $C_6$ are the number of ratings provided by disadvantaged and advantaged users, respectively. 
To be specific, 
$C_5 = \sum\nolimits_{u \in \mathcal{D}}\sum\nolimits_{i \in \mathcal{I}}\mathds{1}_{(u,i) \in \Omega }$, where $\mathds{1}_{(u,i) \in \Omega }=1$ if $(u,i) \in \Omega$ and 0 otherwise.

\section{Additional Results for Synthetic Datasets} \label{appendix_Synthetic}

\begin{table*}[!b]
	\centering
	\addtolength{\tabcolsep}{-0.55pt}
	\setlength{\doublerulesep}{3\arrayrulewidth}
	\caption{Unfairness scores under different methods on synthetic dataset. 
    The synthetic dataset exhibits no population imbalance ($\alpha_1=\alpha_2=0.4$) and has observation bias ($\beta_1=0.2$, $\beta_2=0.05$).
	}
	\label{results_Synthetic_two}
	\footnotesize
		\begin{tabular}{l|l|cccc|cccc|cccc|cccc}
			\toprule
			\multirow{3}[6]{*}{Datasets} & \multirow{3}[6]{*}{Method} & \multicolumn{16}{c}{Fraction of antidote users} 
			\bigstrut\\
			\cline{3-18}           &
			&  \multicolumn{4}{c|}{Value unfairness}      
			& \multicolumn{4}{c|}{Absolute unfairness}        
			&  \multicolumn{4}{c|} {Overestimation unfairness}       
			& \multicolumn{4}{c}{Non-parity unfairness} 
			\bigstrut \\
			\cline{3-18}          &       & 0.5\%     & 1\%     & 2\%    & 3\%    & 0.5\%     & 1\%     & 2\%    & 3\%    & 0.5\%     & 1\%     & 2\%    & 3\%    & 0.5\%     & 1\%       & 2\%    & 3\% 
			\bigstrut \\
			\hline
			\hline
			\multirow{7}[2]{*}{Synthetic} 
			& None &  0.322    &  0.322 &  0.322 &  0.322 & 0.195& 0.195 & 0.195  &  0.195 & 0.161 & 0.161  & 0.161&  0.161 &0.041 & 0.041 & 0.041&0.041 \\
			& Regularization &  0.294    & 0.294 & 0.294 &0.294  & 0.187    &  0.187  &   0.187   &  0.187    &   0.150   &   0.150    &   0.150   &   0.150   &  0.037   &   0.037   &  0.037   & 0.037  \\
			& Maximum &  0.322  & 0.319 & 0.323 &0.316 & 0.189  & 0.191  &  0.186   &  0.194 &  0.159   &  0.159    & 0.160  &  0.156  & 0.042  & 0.039  & 0.050 & 0.046\\
			& Minimum & 0.319  & 0.315 &0.318 & 0.316 & 0.192  & 0.192  & 0.190 &  0.188  & 0.152 &  0.155  & 0.151 &   0.149  & 0.044& 0.049  &0.058 &  0.048\\
			& Random &  0.313  & 0.322 & 0.319 & 0.314 &  0.189  &  0.197  & 0.195   &  0.189  &  0.160  & 0.163  & 0.163  & 0.154  & 0.039  &   0.040  &  0.041&  0.044\\
			& Rastegarpanah et al. & 0.308  & 0.302 &  0.294 & 0.289 &  0.189  & 0.185  & 0.185  &  0.184  &  0.155 & 0.154  &  0.147 & 0.143 & 0.037 &  0.035  & 0.035   &  0.035 \\
			& FairRoad  & \textbf{0.278} & \textbf{0.245} & \textbf{0.215} & \textbf{0.206} &  \textbf{0.184} &  \textbf{0.175} & \textbf{0.173} &\textbf{0.170}  & \textbf{0.146} & \textbf{0.144} & \textbf{0.125} & \textbf{0.110} & \textbf{0.036}& \textbf{0.028} & \textbf{0.027}& \textbf{0.024} \\
			\bottomrule
		\end{tabular}%
	\vspace{2mm}
\end{table*}%

\begin{table*}[!b]
	\centering
	\addtolength{\tabcolsep}{-0.55pt}
	\setlength{\doublerulesep}{3\arrayrulewidth}
	\caption{Unfairness scores under different methods on synthetic dataset. 
    The synthetic dataset exhibits population imbalance ($\alpha_1=0.4, \alpha_2=0.2$) and has no observation bias ($\beta_1=\beta_2=0.2$).
	}
	\label{results_Synthetic_three}
	\footnotesize
	\begin{tabular}{l|l|cccc|cccc|cccc|cccc}
		\toprule
		\multirow{3}[6]{*}{Datasets} & \multirow{3}[6]{*}{Method} & \multicolumn{16}{c}{Fraction of antidote users} 
		\bigstrut\\
		\cline{3-18}           &
		&  \multicolumn{4}{c|}{Value unfairness}      
		& \multicolumn{4}{c|}{Absolute unfairness}        
		&  \multicolumn{4}{c|} {Overestimation unfairness}       
		& \multicolumn{4}{c}{Non-parity unfairness} 
		\bigstrut \\
		\cline{3-18}          &       & 0.5\%     & 1\%     & 2\%    & 3\%    & 0.5\%     & 1\%     & 2\%    & 3\%    & 0.5\%     & 1\%     & 2\%    & 3\%    & 0.5\%     & 1\%       & 2\%    & 3\% 
		\bigstrut \\
		\hline
		\hline
		\multirow{7}[2]{*}{Synthetic} 
		& None &  0.106   & 0.106& 0.106 &0.106  & 0.030  &  0.030 &  0.030  &  0.030  & 0.053 &  0.053   &  0.053  &  0.053  &  0.044  &  0.044  & 0.044  & 0.044  \\
		& Regularization &  \textbf{0.093}    & 0.093 & 0.093 &0.093  & 0.025    &  0.025  &   0.025   &  0.025    &   0.044   &   0.044    &   0.044    &   0.044   &  \textbf{0.041}   &   0.041   &  0.041   & 0.041  \\
		& Maximum &  0.108  & 0.125 & 0.109 &0.111  & 0.028 & 0.030  &  0.041 &  0.049  & 0.051 & 0.058   &  0.048  &  0.049  &  0.041  &  0.041   &  0.042 & 0.043  \\
		& Minimum & 0.098  &0.096 &  0.094 & 0.092 &  0.031  & 0.029   &   0.034  &  0.046  & 0.049   &  0.048 & 0.050   &  0.045  &  0.043 & 0.042  &  0.045   &  0.047 \\
		& Random &  0.103 &0.103 & 0.103 & 0.092 &  0.028  &0.027   & 0.026    & 0.033  &  0.048   &  0.051  &  0.052  &  0.046  & 0.043  &  0.043   &  0.044  &  0.043 \\
		& Rastegarpanah et al. & 0.105  & 0.105 &  0.098 & 0.092 &  0.028  & 0.028  & 0.026  &  0.026  &  0.050 & 0.049  &  0.046 & 0.043 & \textbf{0.041} &  0.041  & 0.040   &  0.040 \\
		& FairRoad &  0.094 & \textbf{0.086} & \textbf{0.073} & \textbf{0.073} & \textbf{0.021} & \textbf{0.021} & \textbf{0.020} & \textbf{0.020}  & \textbf{0.042} & \textbf{0.040} & \textbf{0.040} & \textbf{0.038} & \textbf{0.041} & \textbf{0.039} & \textbf{0.039} & \textbf{0.038} \\
		\bottomrule
	\end{tabular}%
\vspace{2mm}
\end{table*}%

\begin{table*}[!b]
	\centering
	\addtolength{\tabcolsep}{-0.55pt}
	\setlength{\doublerulesep}{3\arrayrulewidth}
	\caption{Unfairness scores under different methods on synthetic dataset. 
    The synthetic dataset exhibits population imbalance ($\alpha_1=0.4, \alpha_2=0.3$) and has observation bias ($\beta_1=0.2, \beta_2=0.1$).
	}
	\label{results_Synthetic_five}
	\footnotesize
	\begin{tabular}{l|l|cccc|cccc|cccc|cccc}
		\toprule
		\multirow{3}[6]{*}{Datasets} & \multirow{3}[6]{*}{Method} & \multicolumn{16}{c}{Fraction of antidote users} 
		\bigstrut\\
		\cline{3-18}           &
		&  \multicolumn{4}{c|}{Value unfairness}      
		& \multicolumn{4}{c|}{Absolute unfairness}        
		&  \multicolumn{4}{c|} {Overestimation unfairness}       
		& \multicolumn{4}{c}{Non-parity unfairness} 
		\bigstrut \\
		\cline{3-18}          &       & 0.5\%     & 1\%     & 2\%    & 3\%    & 0.5\%     & 1\%     & 2\%    & 3\%    & 0.5\%     & 1\%     & 2\%    & 3\%    & 0.5\%     & 1\%       & 2\%    & 3\% 
		\bigstrut \\
		\hline
		\hline
		\multirow{7}[2]{*}{Synthetic} 
		& None & 0.279   & 0.279 & 0.279 &0.279  &0.101   & 0.101  &   0.101 &  0.101   &  0.162  &  0.162   &   0.162  & 0.162   &  0.045  &  0.045  &  0.045 & 0.045  \\
		& Regularization &  0.241    & 0.241 & 0.241 &0.241  & 0.093    &  0.093  &   0.093   &  0.093    &   0.155   &   0.155    &   0.155    &   0.155   &  0.034   &   0.034   &  0.034   & 0.034  \\
		& Maximum &  0.272  & 0.302 & 0.299 &0.263  & 0.099 &  0.095 &  0.099   & 0.102  & 0.157   &  0.162 &  0.153 &  0.129  &  0.042  &  0.053   &  0.043  & 0.052 \\
		& Minimum & 0.254   & 0.255 & 0.246 &0.289 &  0.096   & 0.089    & 0.097  &  0.104  & 0.154  &  0.145  &  0.143   &   0.151  & 0.042 & 0.047  & 0.052   & 0.053 \\
		& Random &  0.278  & 0.269 &0.276 &0.267 &  0.101   & 0.096   & 0.108 & 0.099 &  0.164 &  0.155   &  0.160 & 0.156   & 0.053  &  0.048  & 0.045 &  0.046 \\
		& Rastegarpanah et al. & 0.257  & 0.256 &  0.248 & 0.230 &  0.101  & 0.099  & 0.099  &  0.092  &  0.161 & 0.154  &  0.147 & 0.125 & 0.037 &  0.034  & 0.031   &  0.029 \\
		& FairRoad & \textbf{0.227} & \textbf{0.210} & \textbf{0.157} & \textbf{0.117} & \textbf{0.091} & \textbf{0.087} & \textbf{0.078} & \textbf{0.074}  & \textbf{0.149} & \textbf{0.134} & \textbf{0.127} & \textbf{0.078} & \textbf{0.031} & \textbf{0.031} &  \textbf{0.027} & \textbf{0.026}\\
		\bottomrule
	\end{tabular}%
\end{table*}%

\end{document}